# Constrained TLBO algorithm for lightweight cable-stiffened scissor-like deployable structures

Soumyajit Manna[1], Arijit Sau[1] and Devesh Punera[2*]

[1]School of Infrastructure, Indian Institute of Technology Bhubaneswar, Khordha, India

[2]Assistant Professor, School of Infrastructure, Indian Institute of Technology Bhubaneswar, Khordha, India

*Corresponding author E-mail: devesh@iitbbs.ac.in

**Abstract**

Present works discusses the efficient structural analysis and weight optimization of the cable-stiffened deployable structures. The stiffening effect of cables is incorporated through a matrix analysis based iterative strategy to identify the active and passive cables. The structural form can be easily deployed to cartesian as well as polar coordinates through the arrangement of duplet members. The large span utility of cable stiffened bar members can pose challenges to the deployability due to increased weight. A novel teaching-learning based optimization (TLBO) algorithm is utilized to optimize the overall weight of the structure through efficient section designs with proper constraint on the yield criteria. The penalty function approach is adopted to identify the unfeasible designs. A number of example cases are analysed and comparison is presented with the existing literature to show the suitability of the proposed approach. Finally, a new form of three-dimensional deployable structure is proposed. It is seen that such deployable structure can be accurately analysed using the iterative matrix analysis approach and efficiently optimized using the present algorithm.

# 1. Introduction:

Deployable structures, finding great utility in storage, transportation and fast-construction, are special type of structures of which the size can be easily reduced by some pre-defined folding mechanism. The application of deployable structure is world-wide and deployability conditions differ due to functional requirements of different deployable forms of designed structures, i.e., flat, barrel vault and spherical etc. [1]. Accordingly, the unit structures as well as arrangements vary for different forms. The simple design of temporary structures [2,3], shelters as well as complex mechanisms of outer space observatory radar and antennas [4,5] is often carried out using the bar and cable members. In addition to bars and cables, membranes and plates are also used for the formation of deployable structure. Often, the combination of two types of elements are also used to form composite foldable structure [3,6]. In number of applications like space antenna, cable is generally used to form cable net for which the topology is being optimised and the scissor structures are generally defined for holding and forming the structures [4].

In simple design, the pantograph type of deployable structures contain duplet forms of scissor members [7]. The duplets are generally mentioned 'scissor-structures' which are of three-types based on the orientation and function of rods, i.e., transitional units, polar units, and angular units [8,9] which are shown in Figure 1. Transitional units are defined in Figure 1(a) where the focal lengths are parallel throughout the whole deployment procedure. Similarly, polar units are shown in Figure 1(b) where the focal lengths never get parallel through the whole deployment procedure except the condition where the rods are simplified into a single line. Angular units are defined in Figure 1(c) where the major structural point is that the units are not straight line. Hence, the focal length could not be parallel. The transitional scissor units will provide a plane surface whereas the polar units help to create the curved geometry with facet like units.

It is generally seen that use of cables in deployable structures avoid excessive deformation and govern the topological profile of deployed shape. The arrangement of cables also re-distribute the member forces. However, very limited studies are available on the use of bar-cable structures. Alongside the general methods used for truss analysis, liner-elastic [10], non-linear [11], static [12], kinematic [13] and dynamic [14] procedures are available for bar-cable structures. The cables associated in the structure are found both discrete and continuous in nature. Different analysis methods are developed to treat discrete [12] and continuous [5,15,16] cables separately. One of the early works by Shan [12] considered the discrete cables between the duplet and bar like members. The matrix method of linear analysis is used to identify the member forces. Based on the iterative scheme for identifying cable members under tension only, active and passive cables are located. The arrangement of active and passive cables is governed by the loading and boundary conditions. Unlike regular finite element approach, the uniplets, consisted with three nodes, are considered as a single bar for the analysis. The idea proposed by You and Pellegrino [17] is based on pre-tensioned cables where flexible pantograph can be transformed into a stiff lattice mast. When the pantograph is folded these cables become passive, and hence the structure is still a single degree-of-freedom mechanism that packages into a small volume. In the extended configuration, however, any required stiffness can be achieved by appropriately sizing the cross sections of the cables and of the members of the pantograph.

It is often argued that scissor-like deployable structures, without any form of pre-tensioning in cables, remain less structurally efficient and undergo relatively large deformations. Various forms of tension-strut structures are proposed to overcome such limitations [18,19]. Still, loss of tension and effective connection design remain some of the challenges in such members. In such situation, a simple structural analysis strategy and the improved optimization algorithms can bring a lot of utility to scissor like structures with cables. While the present study does not explicitly identify the connection design, different forms of concentric and eccentric connection designs can be followed from Liew et al. [19].

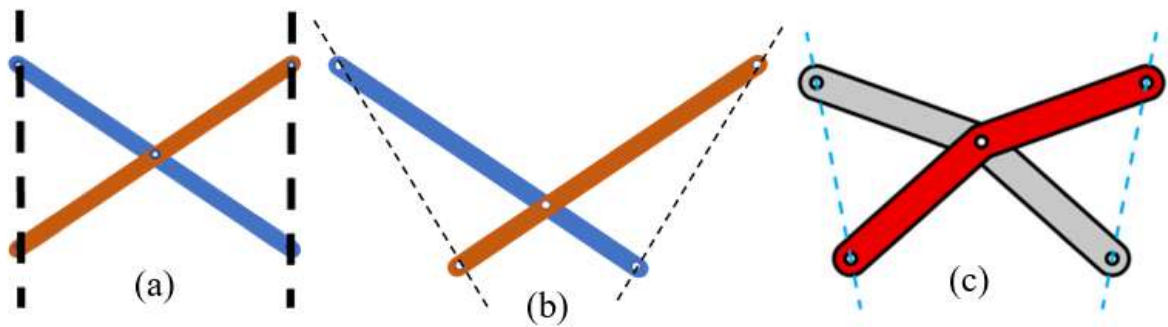

Fig. 1. Different forms of Scissors like Structures. (a) Transitional Units. (b) Polar Units. (c) Angular Units

The optimization of the fixed form trusses and other types of structures are well-established. A few studies on topology optimization of the cable-net [20] are also attempted. In case of trusses, generally population-based metaheuristic optimization techniques are adopted. Among the adopted techniques, Genetic Algorithm (GA) [21–23], Collision based optimization (CBO) [1]

, Ant colony optimization (ACO) [24], Teaching Learning based optimization (TLBO) [25–27], Vibrating particle system (VPS) etc. are the well-known ones. Though GA, ACO, CBO etc. are popularly used for structural optimization, the algorithm specific parameters make the convergence slow. The algorithms require significant population size with large number of iterations. The number of parameters like crossover and mutation used in GA often remain less probabilistic. To overcome this limitation, Rao [25,27,28] introduced the TLBO algorithm which does not have any algorithm specific parameters. The TLBO algorithm gives advantage of faster convergence with lesser population size. The concept of TLBO was introduced based on the learning effects by teachers and students inside classroom. It is being introduced to solve constrained mechanical optimization problems [27]. Being nature inspired, it is also works on population-based method. To the best of the authors knowledge, no optimization studies using novel TLBO algorithm are available for bar-cable deployable structures.

Throughout the existing literature, it is found that there is more focus on the structure with rigid members but not the cables. The application of cables was limited to the formation of cable net. The optimization concept is applied for form finding, i.e., topology optimization with no studies based on efficient TLBO algorithm. Cables are very useful to those structures specifically where the requirements are stiffness but not the strength. This the primary criteria for the space structure as well. The use of cables helps the structure to deploy and retract easily. This work implements an efficient structural analysis strategy for bar-cable based deployable structures. The effective section configuration and sizes are optimized based on novel TLBO algorithm. The results are compared for some classical forms of deployable systems and a new structural form is presented for deployability in curvilinear direction.

## 2. Structural Analysis of bar-cable system

We present an efficient structural analysis strategy for bar-cable based deployable structures in this section. The matrix analysis-based scheme identifies the active and passive cables in iterative manner to develop the overall structural stiffness. The method is based on the earlier work of Shan [12] and evaluates the stiffness matrix for uniplet members with only translational degrees of freedom.

### 2.1 Development of Uniplet stiffness matrix

Stiffness matrix method is a powerful technique for the structural analysis of the skeletal structures. The uniplet member is defined as a bar member having three-nodes with intermediate node as the pivot for the second uniplet member (see Fig. 1a). Such set of two members are called as duplet which is the fundamental feature of many scissor-type deployable structures. A uniplet member can be identified as single element having three nodes with five-degrees of freedom (DoF) at each node excluding the twisting deformation.

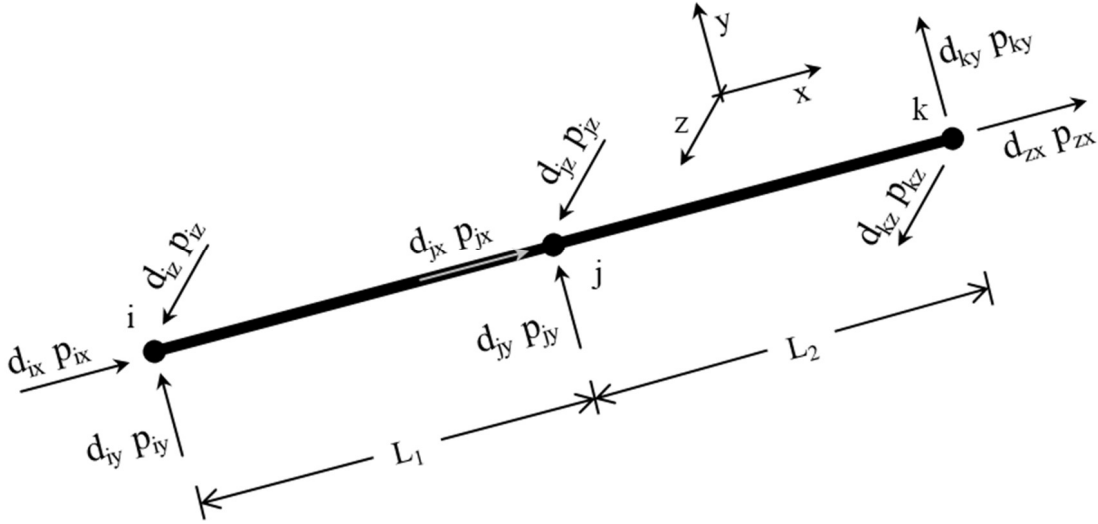

Fig. 2. Arrangement of a uniplet member

The standard linear stiffness for such fifteen-DoF system can be obtained using regular stiffness formulation and can be given in standard convention as;

$$\begin{bmatrix} \mathbf{P} \\ \mathbf{M} \end{bmatrix} = \begin{bmatrix} \mathbf{K}_{DD} & \mathbf{K}_{D\theta} \\ \mathbf{K}_{\theta D} & \mathbf{K}_{\theta\theta} \end{bmatrix} \begin{bmatrix} \mathbf{D} \\ \mathbf{\theta} \end{bmatrix}, \quad (1)$$

here, $\mathbf{P} = \begin{bmatrix} P_{ix} & P_{iy} & P_{iz} & P_{jx} & P_{jy} & P_{jz} & P_{kx} & P_{ky} & P_{kz} \end{bmatrix}^T$ and $\mathbf{M} = \begin{bmatrix} M_{iy} & M_{iz} & M_{jy} & M_{jz} & M_{ky} & M_{kz} \end{bmatrix}^T$ define the force and moment vectors while $\mathbf{D} = \begin{bmatrix} D_{ix} & D_{iy} & D_{iz} & D_{jx} & D_{jy} & D_{jz} & D_{kx} & D_{ky} & D_{kz} \end{bmatrix}^T$ and $\mathbf{\theta} = \begin{bmatrix} \theta_{iy} & \theta_{iz} & \theta_{jy} & \theta_{jz} & \theta_{ky} & \theta_{kz} \end{bmatrix}^T$ describe corresponding translation and rotation DoFs. The uniplet joints are considered without moment loading (**M**=null vector) so rotational degrees of freedom can be defined in terms of translational DoF only and condensed matrix can be obtained for only joint translations as;

$$\mathbf{P} = (\mathbf{K}_{DD} - \mathbf{K}_{D\theta} \mathbf{K}_{\theta\theta}^{-1} \mathbf{K}_{\theta D}) \mathbf{D} \quad (2)$$

The coefficients of the condensed matrix are evaluated numerically in terms of the geometrical and material parameters and same is added in Appendix A. The stiffness matrix for the simple bar members can be adopted same as the two-noded space truss member. The inclusion of the cable stiffness is made through step-wise scheme, explained in the next section.

## 2.2 Analysis of cables:

The cables primarily contribute to the stiffness of the overall structure and regulate/maintain the deployment. Kwan et al. [13] proposed a methodology with pre-determination of active and passive cables having possibility of deployment through the cable pretension. As per them [13,17], the active cables remain taut and get shortened during deployment. Their shortening causes the deployment in the structural form. On the other hand, passive cables generally remain slack during deployment and become taut once the deployed configuration is attained. Present work assumes the cables to be discontinuous between joints with no pretension available. We adopt the method described by Shan [12] which is simpler with lesser complexity, less computational cost and the accurate solution providing efficiency for discrete cables.

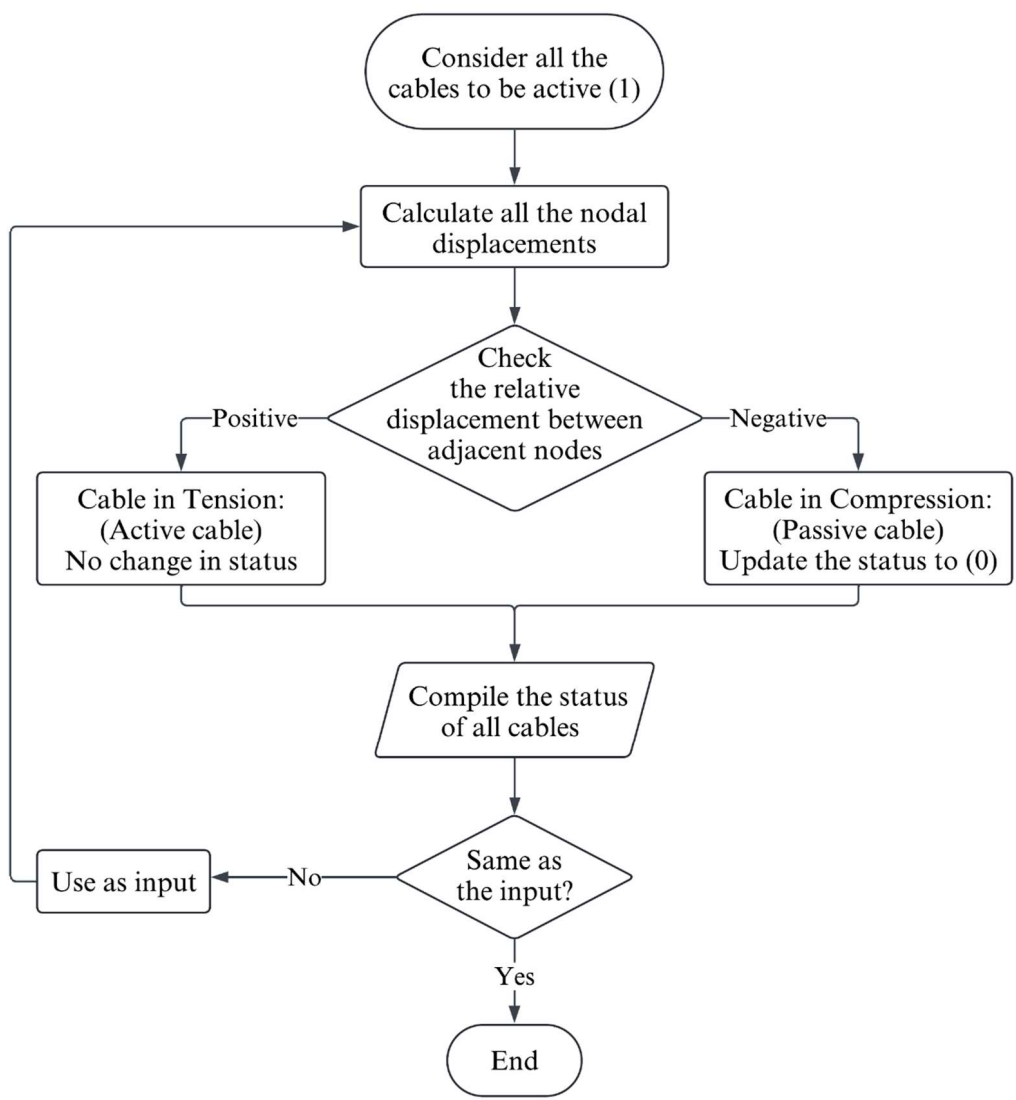

Fig. 3. Flow chart of iterative algorithm

It is postulated that under the external loading criteria, not all cables can remain under tension. An iterative scheme identifies cables subjected to tension (further called as active cables) and cables under compression (called as passive cables) based on the matrix analysis of overall structures. The active cable stiffness is defined similar to bar member stiffness. The iterative procedure begins with considering all cables active and evaluates the joint deformations. The nature of member axial force is located to update the nomenclature of active and passive cables. The iterative scheme stops with convergence of any two subsequent steps for cable nomenclature. Fig. 3 illustrates the overall algorithm flow-chart.

## 3. Deployability Criteria:

In a polar unit, scissor hinges of the system generate a curvature during deployment due to the place of the hinge which creates two unequal arms. The top and bottom hinges and the scissor hinges lie on concentric circles. The deployability of the structure is followed by the basic geometric principles of concentric circles. The ratio of both the arms in each uniplet should be constant to provide the proper concentric circular form. The deployability condition is stated as:

$$\frac{a_{i-1}}{b_{i-1}} = \frac{a_i}{b_i} = \frac{a_{i+1}}{b_{i+1}} = \cdots = \frac{a_n}{b_n} \qquad (3)$$

here, $a_i$ and $b_i$ are the semi bars of the $i^{th}$ uniplet. For reference, $a_i$ and $b_i$ are equivalent to $l_1$ and $l_2$ as mentioned in Fig. 4. The angle swept at centre is defined by number of duplet units and total central angle of each unit arch. The relationship between number of unit cells and the angle of each unit cell is given as:

$$N = \frac{\alpha}{\varphi} \qquad (4)$$

here, N defines the number of unit cells, $\alpha$ is the total angle covered by the curvilinear structure and $\varphi$ is the angle covered by each unit cell. The relation of the radii of both imaginary concentric circles can be defined in terms of the outer radius R, inner radius $R_b$ and deployed width along the radial direction t.

$$R = R_b + t \qquad (5)$$

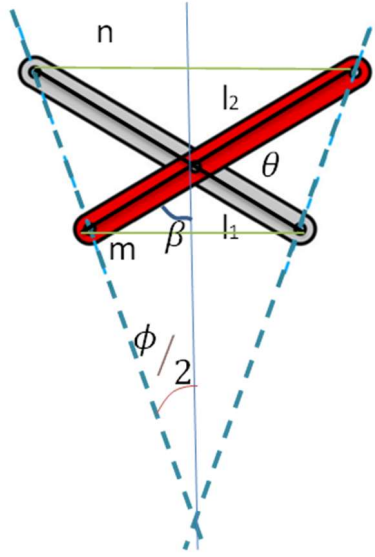

Fig. 4. Unit cell of the selected structure

Polar units have identical bars of span L which are divided into two unequal semi-bars, $l_1$ and $l_2$. The length L can be defined using the cosine rule as:

$$L^2 = R_b^2 + R^2 - 2RR_b \cos\phi \tag{6}$$

Further, with the similarity of the triangles,

$$\frac{R_b}{R} = \frac{m}{n} = \frac{l_2}{l_1} \tag{7}$$

here, m/n is pre-defined fixed ratio as per the requirement and usability of the structure.

So, the values of both the unequal semi bars are,

$$l_1 = \frac{LR}{R+R_b}, \qquad l_2 = \frac{LR_b}{R+R_b} \tag{8}$$

And value of β is obtained as,

$$\beta = 2\sin^{-1}\left[\frac{R_b}{l_2}\sin\frac{\phi}{2}\right] \tag{9}$$

On the basis of these parameters a deployable structure is proposed having the possibility of polar coordinate deployment. The inner radius is taken as 3 m while the radial width is 0.45 m with α = 180°, ϕ = 36°. The structure selection is dependent on the architectural feasibility of structure and it should be kinematically satisfied. The one more criterion had to be followed here for the design of the structure. It is the cable arrangements. The algorithm mentioned in the manuscript [12] is based on discrete cable. The proposed structural form is shown in Fig. 5. The solid lines represent duplet members while the dashed lines show cables.

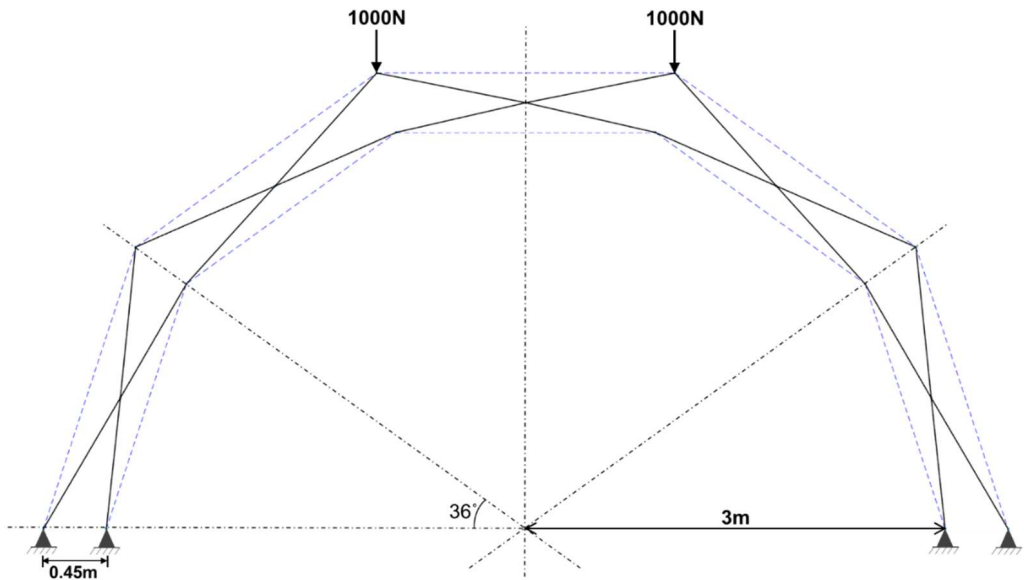

Fig. 5. Selected Structure for analysis

## 4. Teaching learning based-optimization (TLBO) algorithm

The structural weight minimization is done based on the teaching learning based optimization algorithm. TLBO is a metaheuristic algorithm inspired by the teaching and learning process in a classroom [25]. It is used to solve optimization problems that are difficult or complex to solve using traditional methods. The basic idea behind TLBO is to simulate the teaching and learning process in a classroom, where a teacher helps students learn by sharing knowledge and experience. In TLBO, the problem is represented in terms of given control variables for a population of solutions. Any $i^{th}$ population can be given in terms of the $n$ control variables as, $\mathbf{X}^i = [x_1^i, x_2^i, x_3^i, ........x_n^i]$. The best solution for the given population set is defined as $\mathbf{X}^{teacher}$. The mean of the individual variable $j$ is defined as $x_j^{mean} = \frac{1}{NP}\sum_{i=1}^{NP} x_j^i$. The estimates of each population set are improved using the difference of the best solution and the mean solution. The new improved solution for $i^{th}$ population is given as,

$$\mathbf{X}_{new}^i = \mathbf{X}_{old}^i + r\left(\mathbf{X}^{teacher} - TF \times \mathbf{X}^{mean}\right) \qquad (10)$$

here, r represents a random number between [0 1] and TF is teaching factor generally taken as 1 or 2. Further, the different populations (other than the teacher phase) interact with each other and improve individual scores. This phase is known as the learner phase. The learning process involves two steps as well; first, each student compares their solution to their classmates' solutions, and if they find a better solution, they adopt it. Second, each student generates a new solution based on their own knowledge and compares it to their current solution. If the new solution is better, it replaces the old solution. If a given population $p$ is better than $q$ the updated population is defined as,

$$\mathbf{X}_{new}^{p} = \mathbf{X}_{old}^{p} + r\left(\mathbf{X}_{old}^{p} - \mathbf{X}_{old}^{q}\right), \quad \text{Otherwise} \quad \mathbf{X}_{new}^{p} = \mathbf{X}_{old}^{p} + r\left(\mathbf{X}_{old}^{q} - \mathbf{X}_{old}^{p}\right) \quad (11)$$

Over time, this process leads to the formation of a population of high-quality solutions that converge towards the optimal solution. TLBO has been successfully applied to a wide range of optimization problems, including engineering design, image processing, and financial forecasting. Overall, TLBO is a powerful optimization technique that can find high-quality solutions to complex problems in a relatively short amount of time. Its ability to balance exploration and exploitation and adapt to changing environments makes it a popular choice for many researchers and practitioners. A simple flow chart of the algorithm is defined in Fig. 6.

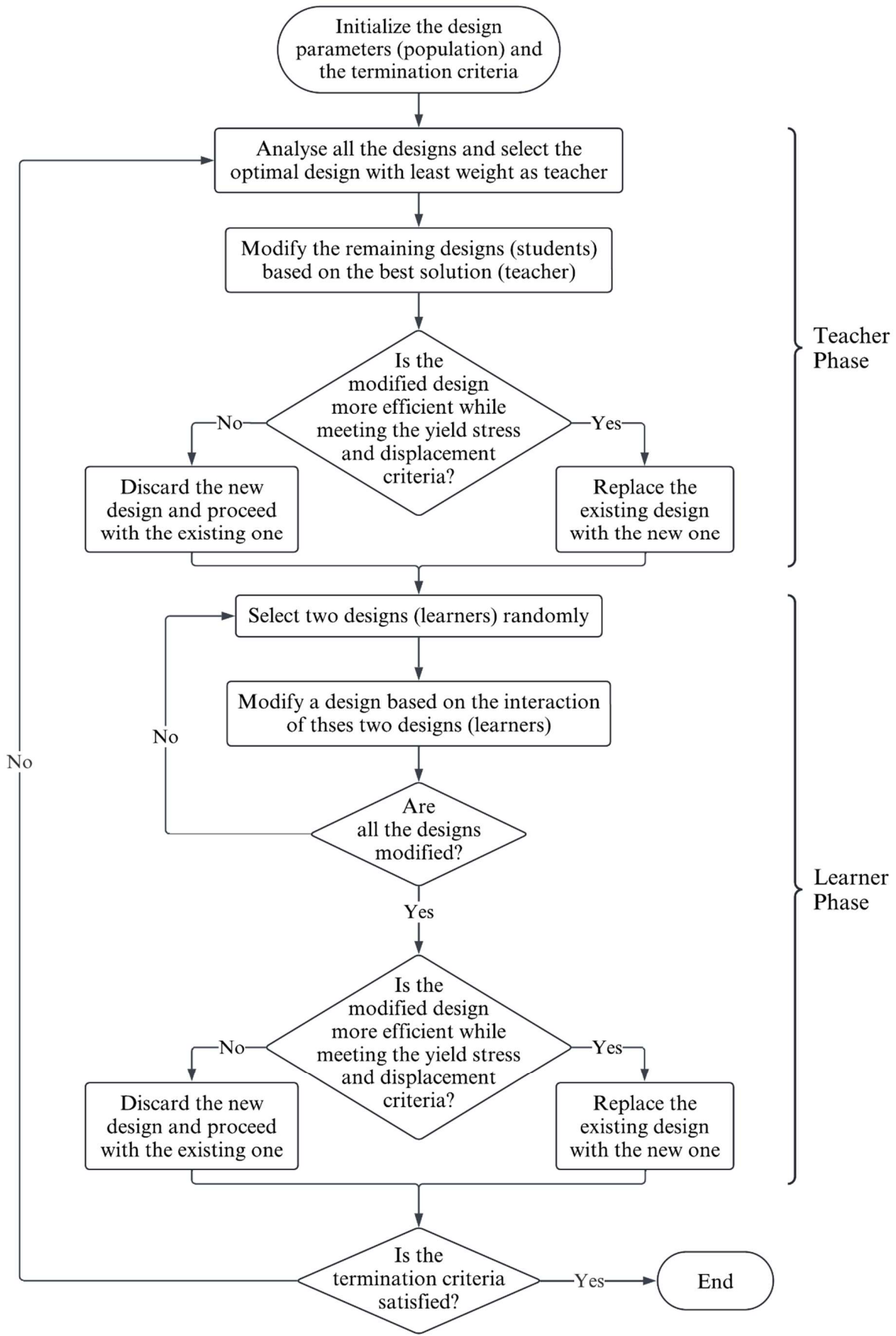

Fig. 6. Flow chart of Teaching Learning Based Optimization Algorithm

## 4.1 Objective function and constraint conditions:

The objective of the present optimization problem is to minimize the weight of the structure while the cable-based structure geometry is fixed. The condition of member yield strength and maximum allowable displacement at the nodal points is also defined as the constraint conditions in the algorithm. A set of area (based on available commercial sizes) are defined for possible selection as the member cross-section in the optimization process.

The objective function can be expressed as,

$$w = \sum_{e=1}^{N_m} \gamma_e L_e A_e \quad \text{with} \quad A^L \leq A_e \leq A^U \tag{12}$$

The constraint conditions are defined as,

$$\sigma_e \leq \sigma^y \tag{13}$$

$$\delta_c \leq \delta^{max} \tag{14}$$

where $w$ is the weight of the cable-based structure composed of $N_m$ members and for each member e, $\gamma_e$ is the material unit weight, $L_e$ is the length, and $A_e$ is the cross-sectional area of the members. The truss design must satisfy limits on member stresses $\sigma_e$ and deflections $\delta_c$ at each node c. Limits on cross-sectional area are given by values defined at lower $L$ and upper $U$ boundaries whereas maximum member stress and node deflection are limited to yield stress $\sigma^y$ and maximum permissible deflection $\delta^{max}$, respectively.

A penalty function is used to account for infeasible structural designs. In this formulation, the value of the objective function (the structural weight) is multiplied by a cumulative penalty that is proportional to the amount of any stress and deflection constraint violations [29].

$$\text{If } \sigma_e \leq \sigma^y, \text{ then } \emptyset_\sigma^e = 0$$

$$\text{If } \sigma_e > \sigma^y, \text{ then } \emptyset_\sigma^e = \left|\frac{\sigma_e - \sigma^y}{\sigma^y}\right|$$

The total stress penalty is defined as:

$$\emptyset_\sigma^k = \sum_{e=1}^{N_m} \emptyset_\sigma^e \tag{15}$$

The penalties for the deflections in x, y and z directions are defined as for each node c,

$$\text{If } \delta_{c(x,y,z)} \leq \delta^{max}, \text{ then } \emptyset_{\delta(x,y,z)}^c = 0$$

$$\text{If } \delta_{c(x,y,z)} > \delta^{max}, \text{ then } \emptyset_{\delta(x,y,z)}^c = \left|\frac{\delta_{c(x,y,z)} - \delta^{max}}{\delta^{max}}\right|$$

So, total deflection penalty is defined as

$$\emptyset_\delta^k = \sum_{c=1}^{N_c}[\emptyset_{\delta x}^c + \emptyset_{\delta y}^c + \emptyset_{\delta z}^c] \tag{16}$$

The total penalty is defined as

$$\psi^k = \left(1 + \emptyset_\sigma^k + \emptyset_\delta^k\right)^\varepsilon \tag{17}$$

Where $\varepsilon$ is the positive penalty exponent. The value of the penalized fitness function is a product of the design of truss and its total penalty which is defined as, $F^k = \psi^k w^k$

Two parameters are required for the optimization work of the cable-based structure which are required for initialization. Generally adopted values are as following, if not specified in the particular problem.

- Population size: 25 (It is the number of the off-springs generated)
- Number of Iterations: 100

## 5. Numerical Analysis

Based on the developed algorithm, a comprehensive numerical analysis is carried out in this section. First, the validation of the present analysis strategy is done using some existing problems from literature for both two-dimensional and three-dimensional cases. Further results are obtained for proposed new structure and optimization of the different deployable configurations.

### 5.1 Validation for structural analysis procedure

**Problem 1- 2D Problem statement**

The problem illustrated in Fig. 7 is adopted for validation of the present analysis scheme. The structure comprises cables, bars, and duplet members, with fixed supports specified at joints 1, 3, 6, and 7. Additionally, a hinge joint at joint 4 configures it as a duplet system. The solid members define duplet or bar members while the cables are shown with dashed lines. Geometrical details and loading conditions are highlighted Fig. 7. The bars and uniplets share an identical tubular hollow section, characterized by a cross-sectional area of 28 mm² while the cables have a cross-sectional area of 6.28 mm². Uniplets have the second moment of area of 290 mm⁴. The material of all members is assumed to have a modulus of elasticity of 210,000 N/mm².

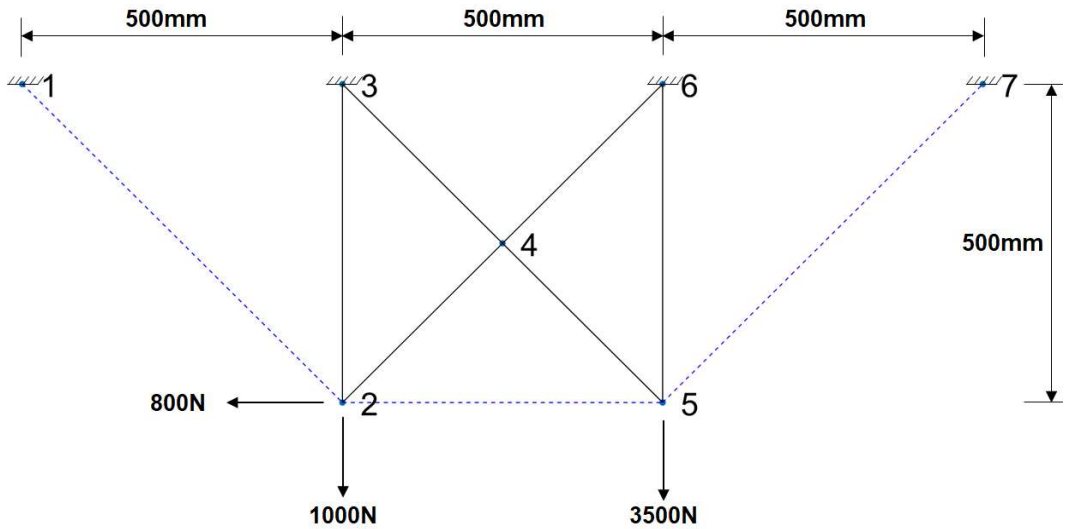

Fig. 7. The 2D validation problem sourced from [12].

Table 1. Displacement and support reaction response for 2D bar-cable structure

| Nodes | Obtained Results (Rounded upto 3 decimals) | | Provided Results [12] | |
|---|---|---|---|---|
| | **Displacements comparison (units are in mm)** | | | |
| | $D_x$ | $D_y$ | $D_x$ | $D_y$ |
| 2 | -0.166 | -0.019 | -0.166 | -0.019 |
| 4 | -0.026 | -0.067 | -0.026 | -0.067 |
| 5 | -0.156 | -0.238 | -0.156 | -0.238 |
| | **Reaction forces comparison (units are in N)** | | | |
| | $R_x$ | $R_y$ | $R_x$ | $R_y$ |
| 1 | 0 | 0 | 0 | 0 |
| 3 | -339.3 | 566.9 | -339.3 | 566.9 |
| 6 | 772.4 | 3566.09 | 772.4 | 3566.2 |
| 7 | 366.9 | 366.9 | 366.9 | 366.9 |

Initially the cables are assumed to be tight and subsequently, the cable status has been evaluated based on the algorithm provided in Section 2.2. It is observed that with the successive iterations, cables converge to a case with two active and one passive member. Table 1 gives the displacement and support reactions of the structural form at different nodes. The obtained results agree well with the existing literature.

Table 2. Member forces comparison (Units are in N)

| Member No. | Obtained Results (Rounded upto 1 decimal) | Provided Results [12] |
|---|---|---|
| 1 | 0 | 0 |
| 2 | 27.6 | 27.6 |
| 3 | 518.9 | 518.9 |
| 4 | 227.6 | 228.1 |
| 5 | 480.0 | 480.2 |
| 6 | 1092.3 | 1092.6 |
| 7 | 2793.7 | 2793.3 |

Table 2 shows the comparison between the obtained results and the provided results for the member forces in local direction. The member forces are rounded up to single decimal place. There are some slight changes in the decimal places which is considered as the rounding off errors on the system. These results validate the accuracy of the proposed algorithm for further use.

**Problem 2- 3D Problem statement-1**

Further, we take up a three-dimensional problem as illustrated in Fig. 8 [30] to validate the usability to the 3D deployable structures. The structure comprises cables, bars, and duplet members, with fixed supports at the four corners. Details regarding the spacing and load combinations of the system are delineated in the Fig. 8. The figure illustrates a space-foldable pantograph structure, referred to as a p-structure, consisting of three types of elements. The dotted lines on the bottom layer represent cables, with four vertical bars located at the supports, and scissor-link units, termed duplets, distributed throughout the structure. The cables, bars, and uniplets share a uniform cross-sectional area of 28 mm². The second moment of area for the uniplet cross-section is defined as 290 mm⁴. The cables are initially taut and unstressed. The material of all members is presumed to possess a modulus of elasticity of 210 GPa.

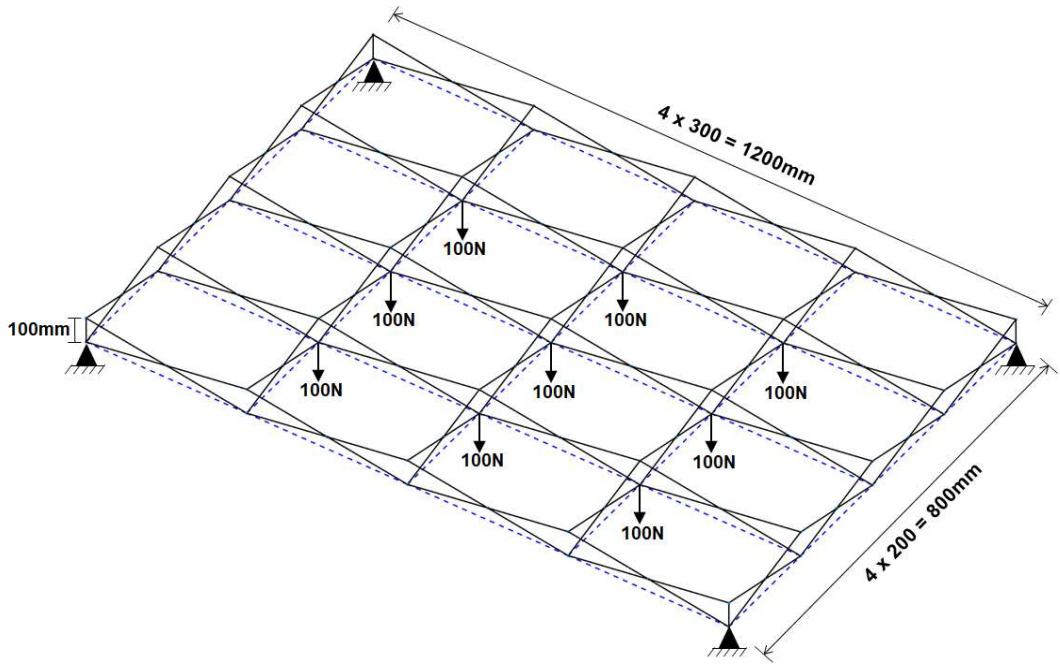

Fig. 8. Schematic of 3D deployable structure for validation [30]

Table 3 compares the results for deflection at the centre point (maximum) of the given structure considering all elements as either uniplet or duplet. The results agree well with the existing results and verify the accuracy of the present analysis approach. Further, Fig. 9 identifies the deflection at different nodes of the bottom layer. The results appear symmetric which is as expected for the symmetric loading and geometry.

**Table 3**. Comparison of max deflection when all the cables are active.

| Type of element | Number of nodes | Provided Max deflection (mm) [30] | Obtained Max deflection (mm) | Status |
|---|---|---|---|---|
| Uniplet | 90 | 12.70 | 12.699 | Result is Validated. |
| Duplet | 50 | 12.70 | 12.699 | Result is Validated. |

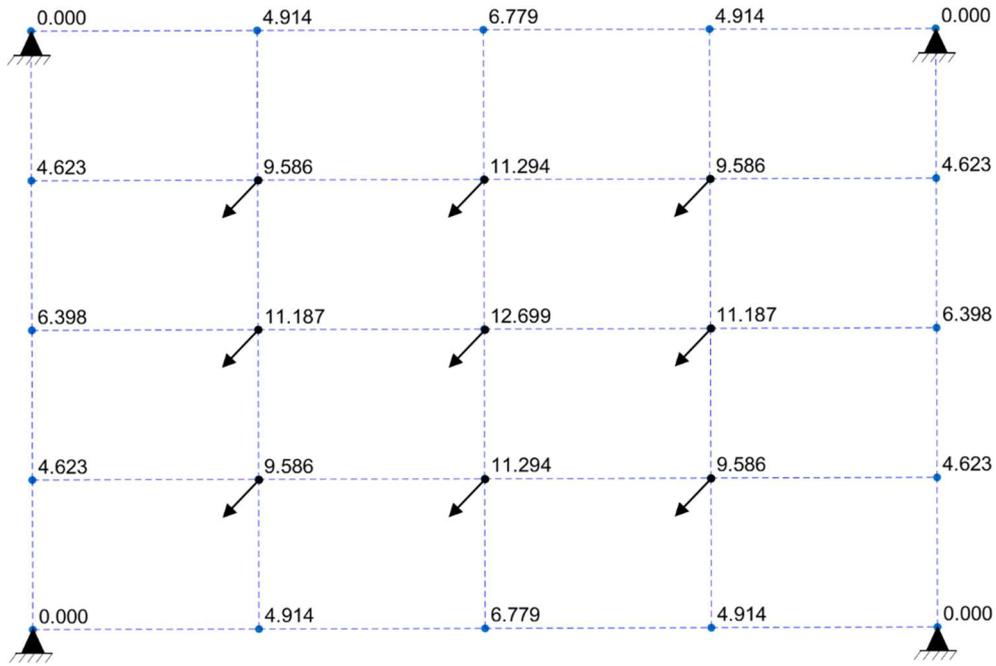

Fig. 9. Transverse displacement at each node in the bottom layer of 3D deployable structure

After validating the results, the problem can be subjected to further iterations to ascertain the status of each cable, determining whether it is active or inactive within the structure. Subsequently, leveraging this information allows for the identification of the minimum number of cables required and their optimal positioning, all without compromising the structural integrity. Fig. 10 highlights the different steps of the iterative procedure to identify the active cables. It is seen that the final converged step has only a few active cables at the bottom layer of the deployable structure.

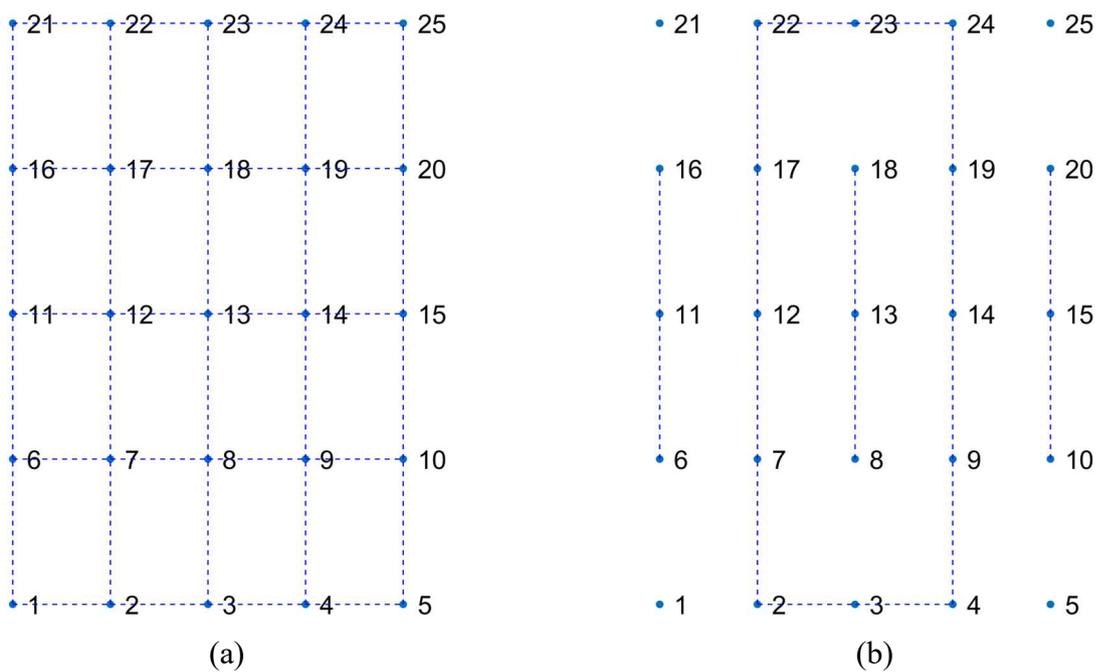

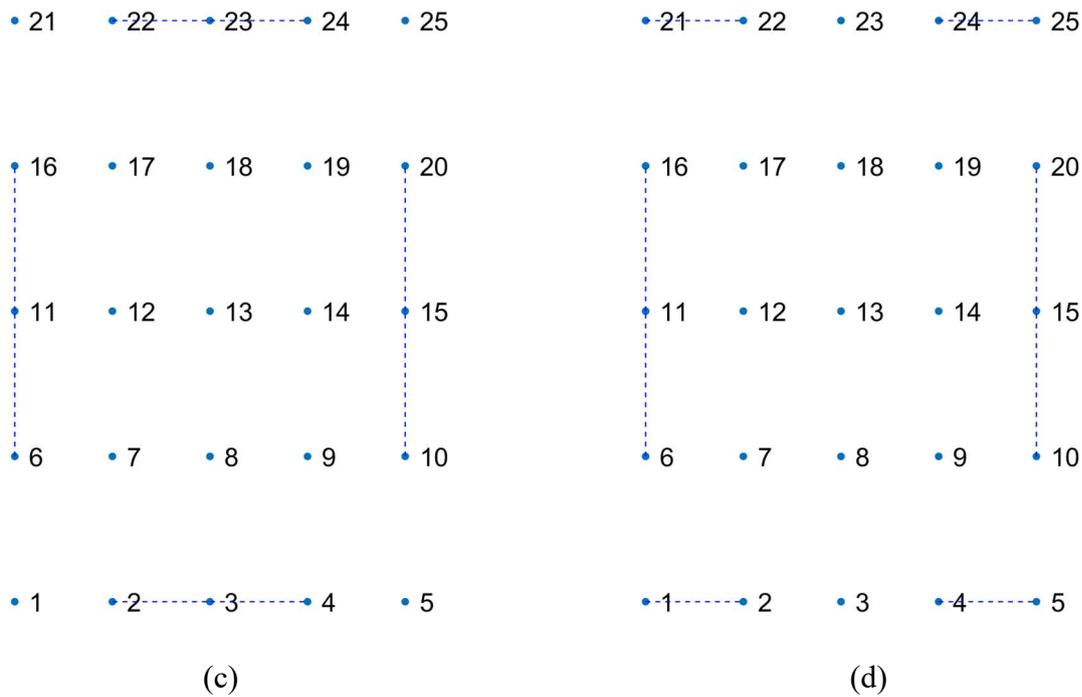

    (c)                                       (d)

Fig. 10. Cable status (a) at the start of analysis, (b) after 1st iteration. (dotted line represent the cable is active) (c) after 2nd iteration, (b) after 3rd and final iteration.

### 5.2. Numerical analysis of the designed deployable structure

**Problem 3- 2D Deployable Structure**

Based on the validation of the structural analysis algorithm, we extend the procedure to analyse the 2D deployable structure as identified in section 4. Fig. 11 illustrates a deployable foldable structure consisting of two types of elements. The blue dotted lines represent cables and the black solid lines represents scissor-link units, termed duplets. Duplets are composed of two uniplet elements. The uniplets have a cross-sectional area of 28 mm². The second moment of area for the uniplet cross-section is specified as 290 mm⁴. The cables, initially taut and unstressed, possess a cross-sectional area of 6.28 mm². The material of all members is assumed to have a modulus of elasticity of 210 GPa.

The algorithm assesses the system and provides information regarding the status of the cables, indicating whether they are in active or passive mode. Fig. 12 shows the status of the active cables after the end of the iterative analysis procedure. It is seen that three cable members remain inactive in the final deployed form under given symmetric loading. The member forces in uniplet members as well as cable members are shown in Figs. 13(a) and (b), respectively.

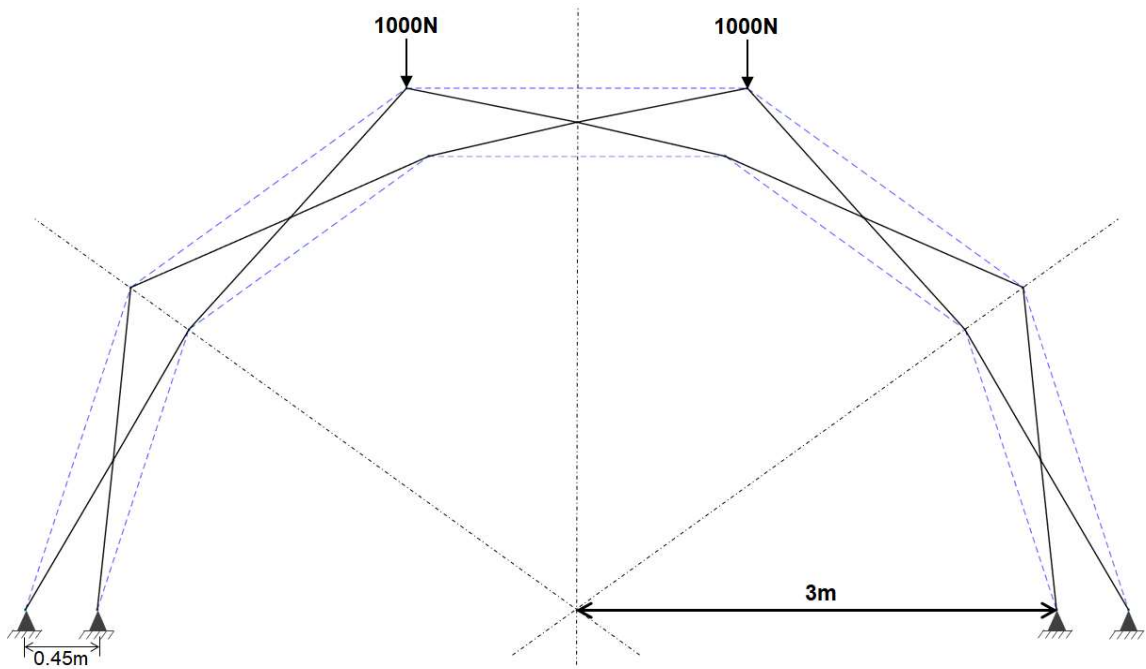

Fig. 11. Vertical symmetrical loading on the 2D foldable structure.

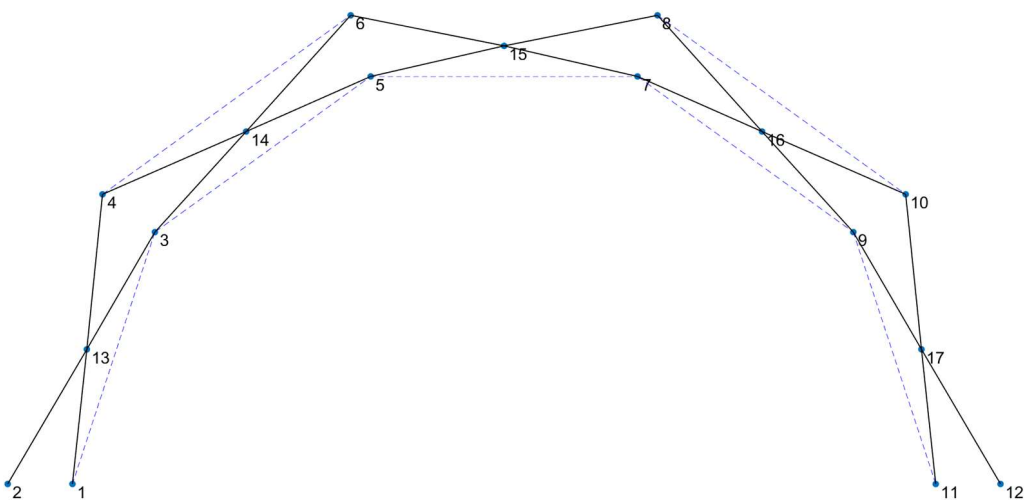

Fig. 12. Cable status after 2$^{nd}$ and final iteration. (dotted line represent the cable is active)

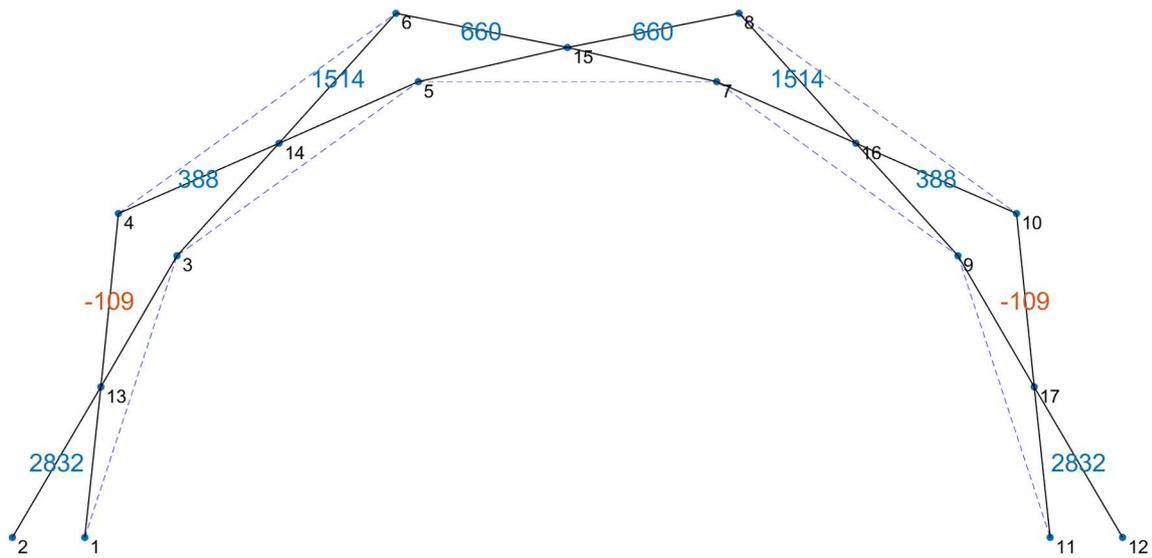

(a)

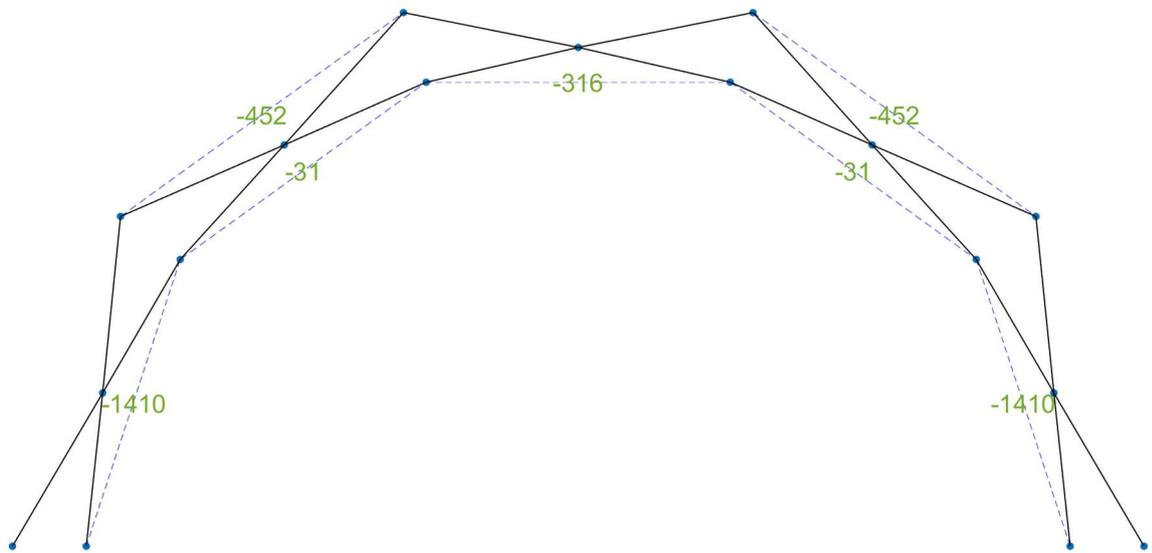

(b)

Fig. 13 (a) Member forces acting at each uniplet in N, (b) Forces acting on each active cable in N. (Negative values represent tension, and positive values represent compression)

**Problem 4- 3D Deployable Structure**

Finally, a 3D deployable structure based on the proposed design is presented. The section can deploy in the polar coordinate with possibility to deploy along the longitudinal axes as well. Fig. 14 illustrates a deployable foldable structure consisting of two types of elements. The blue dotted lines represent cables and the black solid lines represents scissor-link units, termed duplets. The system consists of two 2D deployable structures, identical to those used in problem 3, connected by similar duplet members and cables, as depicted in the figure. The uniplets have a cross-sectional area of 28 mm². The second moment of area for the uniplet cross-section is specified as 290 mm⁴. The cables, initially taut and unstressed, possess a cross-sectional area of 6.28 mm². The material of all members is assumed to have a modulus of elasticity of 210,000 N/mm². A symmetric loading is applied in the transverse direction to selected nodes. The geometrical dimensions are same as described for the proposed geometry in Section 4.

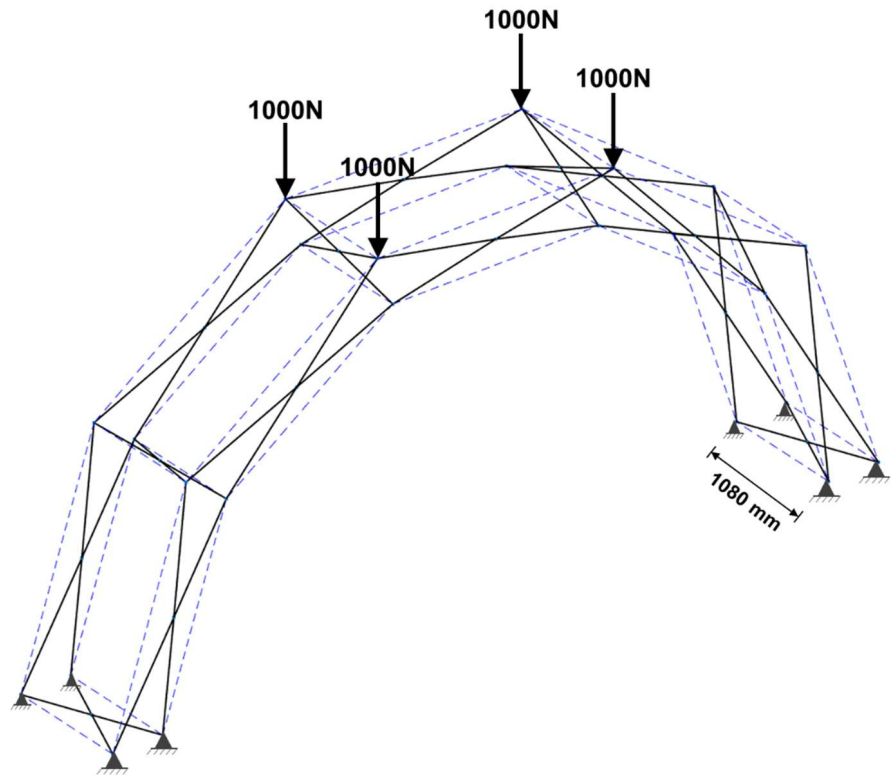

Fig. 14. 3D view of the deployable structure with symmetric vertical loading.

The algorithm assesses the system and provides information regarding the status of the cables, indicating whether they are in active or passive mode. The active cables appear similar to problem 3 in the plan view. Fig. 15(a) shows the structure view with only active cables and duplet members after the completion of the iterative scheme. The only active set of cables are highlighted in Fig. 15(b). Further, Fig. 16 shows the forces in the duplet member at the front and top view schematics. The forces in the active cables are highlighted in Fig. 17 (front view) and Fig. 18 (top view). It is seen that number of cables remain under compression (inactive) while a few duplet members also remain unstressed.

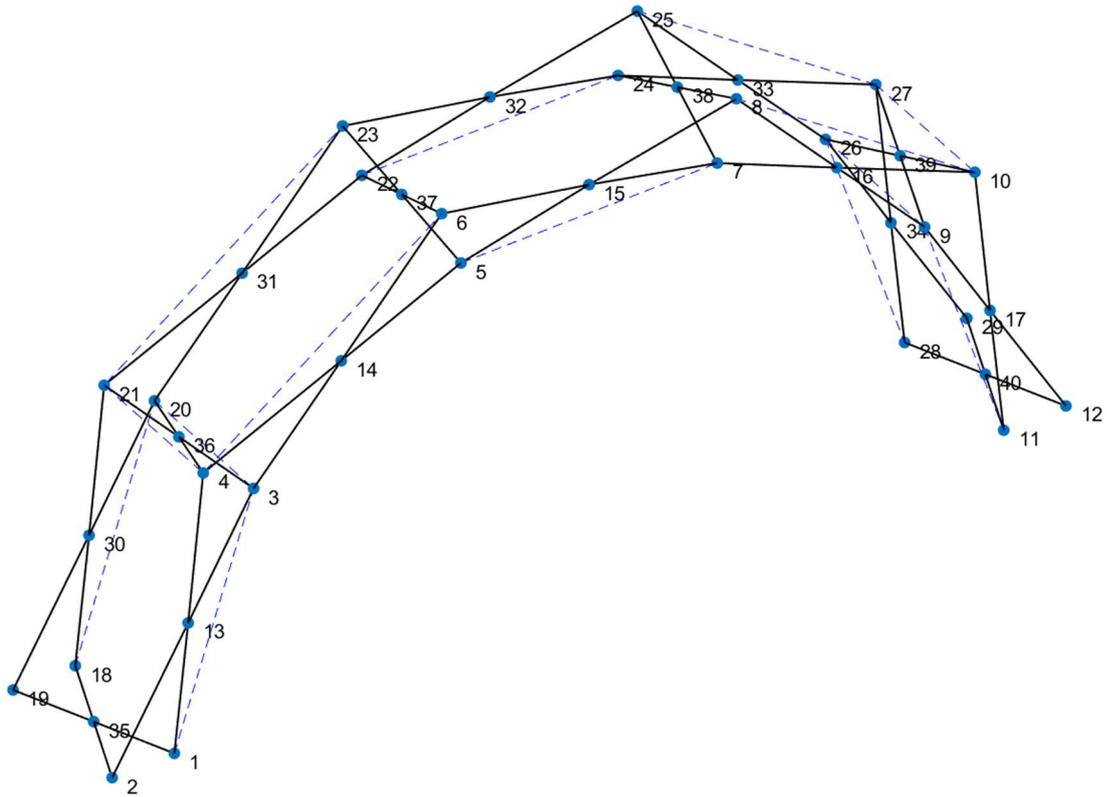

(a)

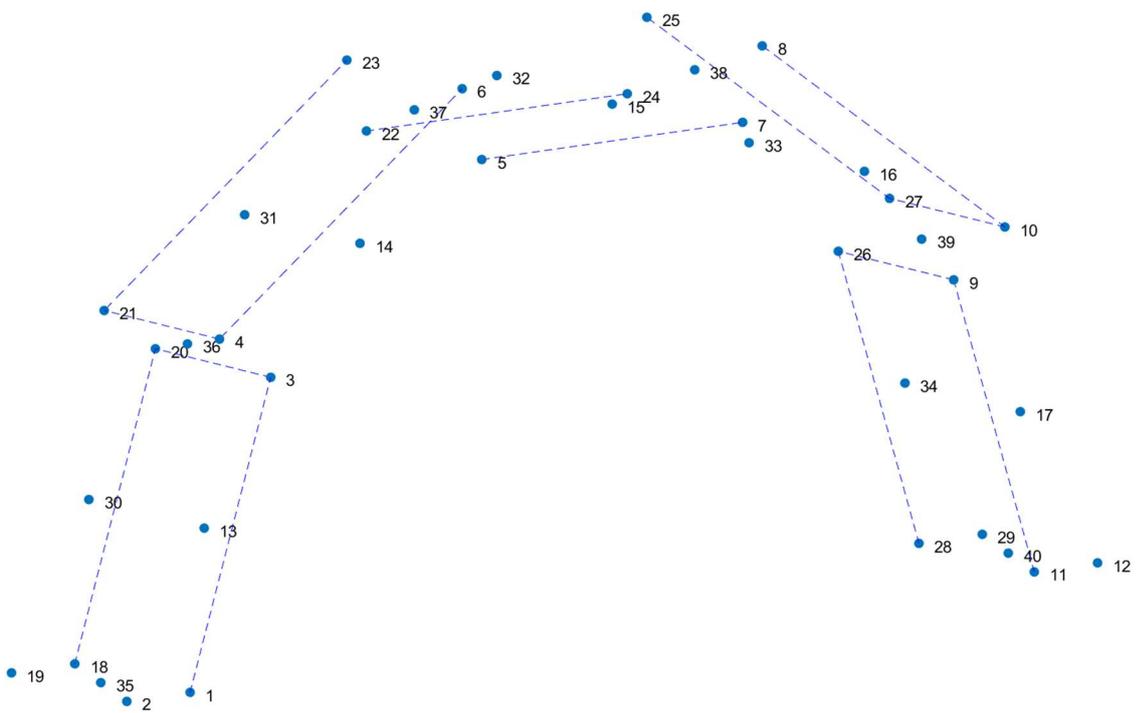

(b)

Fig. 15. Cable status after final iteration (a) 3D view with cables and duplets, (b) 3D view with only cables along with node numbers. (Dotted lines represent active cables)

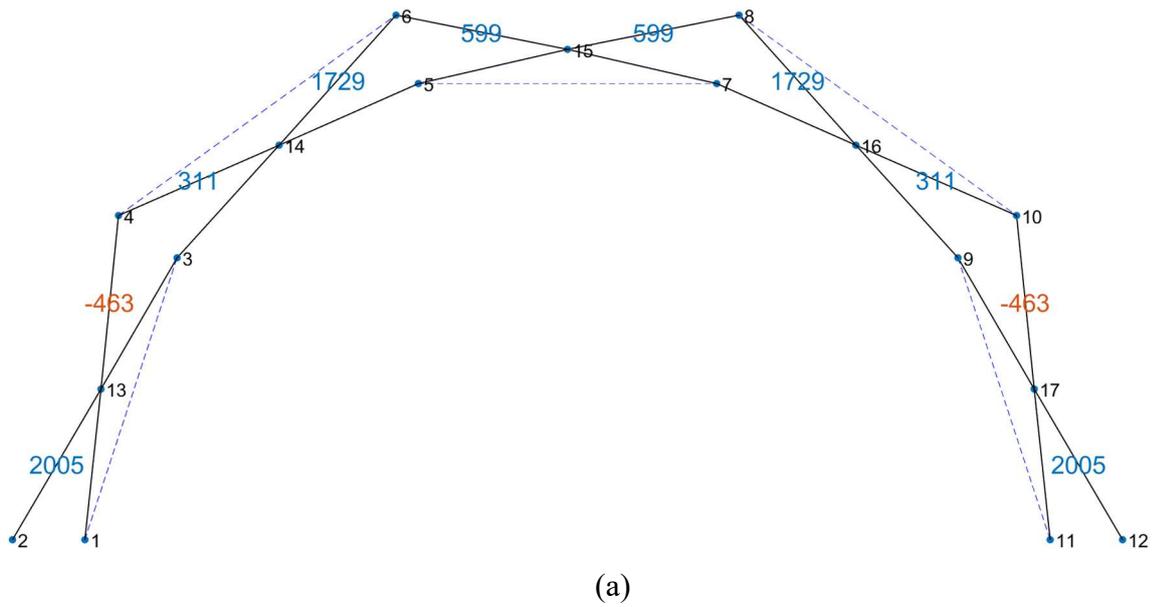

(a)

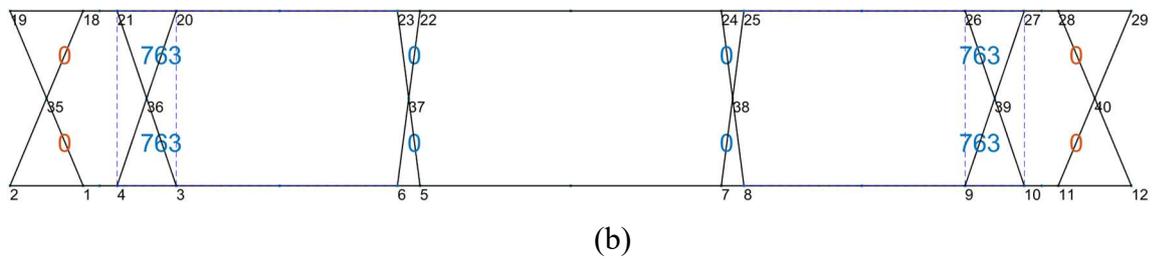

(b)

Fig. 16. Member forces at each uniplet in N, (a) Front view (b) top view. (Negative values represent tension, and positive values represent compression)

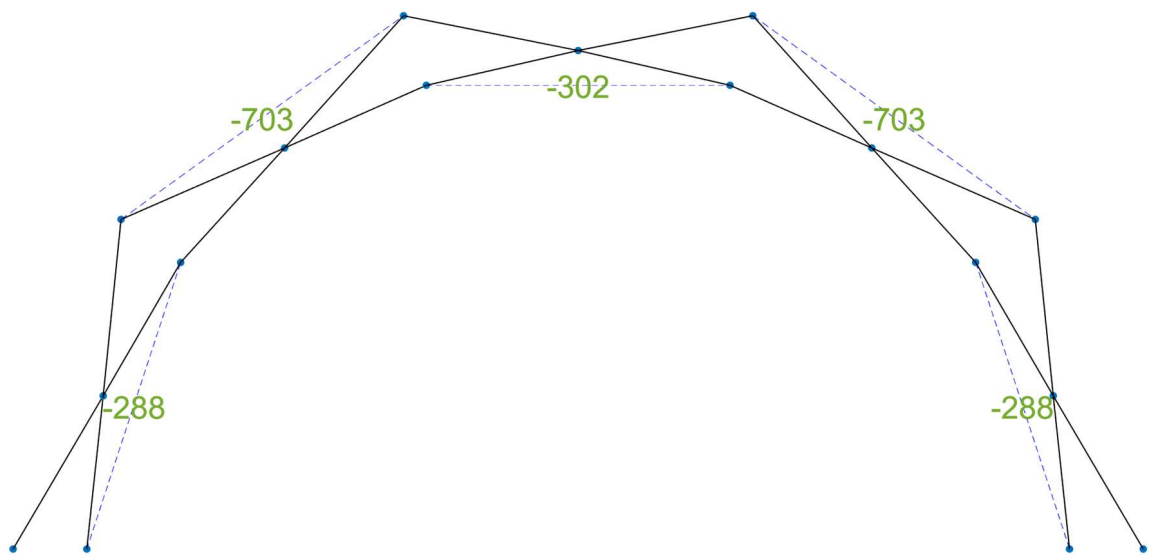

Fig. 17. Forces acting on each active cable (in N) at the front view

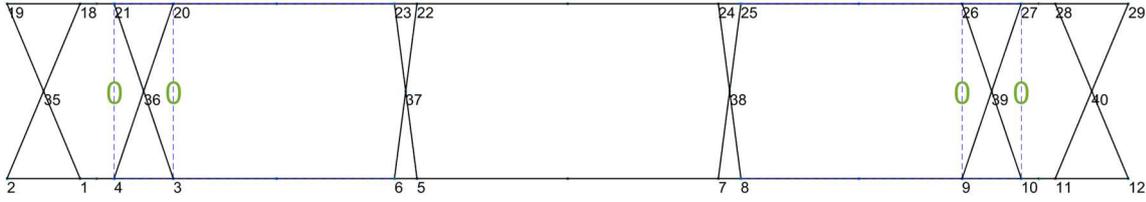

Fig. 18. Forces acting on each active cable (in N) at the top view

### 5.3. Constrained weight optimization using TLBO algorithm

The ease of erection for deployable structures is often dependent on their total weight. While analysing the existing structural configuration, it is seen that the member forces vary significantly with a few members being less stressed whereas a few stressed beyond the yield value. Further, joint deformations at times may not remain small as assumed in the linear theory. An efficient structural optimization can obtain the least weight of the structure while maintaining the stresses and joint deformation within permissible limits. For the weight optimization of the structures, we have considered steel as the material for all the cable, bar, and uniplet members, with a mass density of 7850 kg/m³ and a Young's modulus of 210 GPa. For safe deployment and operation, the allowable stress in the member is considered as 200 MPa, and the allowable nodal displacement is considered as 5 mm (although it will vary depending on the size of the structure).

### 5.3.1. Implementation to 2D Problem

The optimization algorithm is implemented to Problem 1 given in section 5.1 and algorithm specific parameters for the TLBO algorithm are as follows,

- Population size: 50 (The number of solutions evaluated per iteration.)
- Number of Iterations: 100 (The number of iterations to be performed.)

Number of function evaluations represents the total number of solutions evaluated after a certain number of iterations are complete. Fig. 19 shows the convergence of the optimization results for total weight of the structure. Having a small number of members, the problem converged faster with 77% reduction in the initial weight. The optimum member area and forces are shown in Table 4.

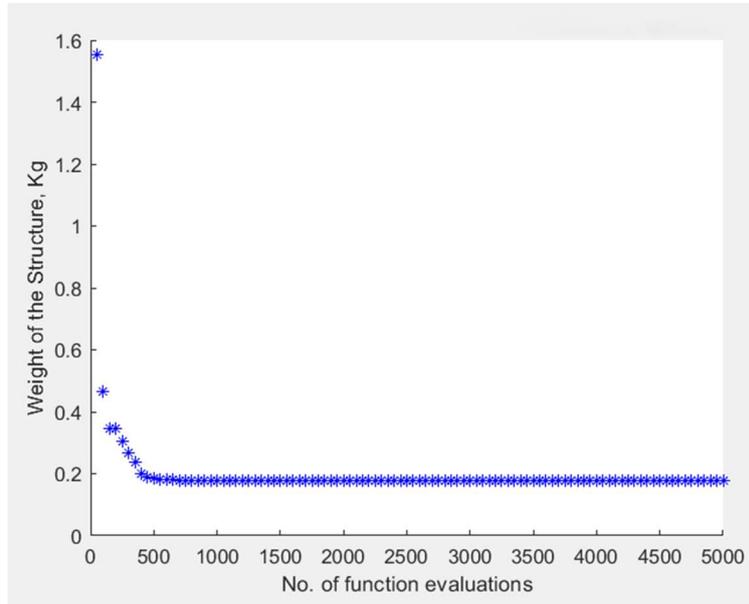

Fig. 19. Convergence graph for the weight optimization on Problem 1

Table 4. Comparison of member force with nearest optimized cross-section area for 2D deployable structure given in problem 1

| Member | Optimized Area | Area Provided | Member Force (Before) | Member Force (After) |
|---|---|---|---|---|
| 1 | 6.28 | 6.28 | 0.00 | 0.00 |
| 2 | 6.28 | 6.28 | -27.61 | -505.05 |
| 3 | 6.28 | 6.28 | -518.95 | -739.54 |
| 4 | 4.41 | 7.07 | -227.62 | -705.05 |
| 5 | 0.785 | 0.785 | -479.90 | -25.29 |
| 6 | 1.07 | 3.14 | -1092.32 | -417.13 |
| 7 | 13.99 | 19.73 | -2793.70 | -2959.18 |

Further, the weight optimization is implemented to Problem 3 as discussed earlier in Section 5.2 with population size of 75 and 200 number of iterations in the TLBO algorithm. The results for the converged solutions are shown in Fig. 20 and optimized area for some of the bar and cable members are shown in Table 5. It is interesting to see that the optimum area is more than the originally selected configuration as the yield stress and permissible deflection limits were not constrained earlier. As the previously provided area is insufficient to support our allowable stress and displacement conditions, we had to significantly increase it based on the requirements, resulting in an increase in the weight of the structure.

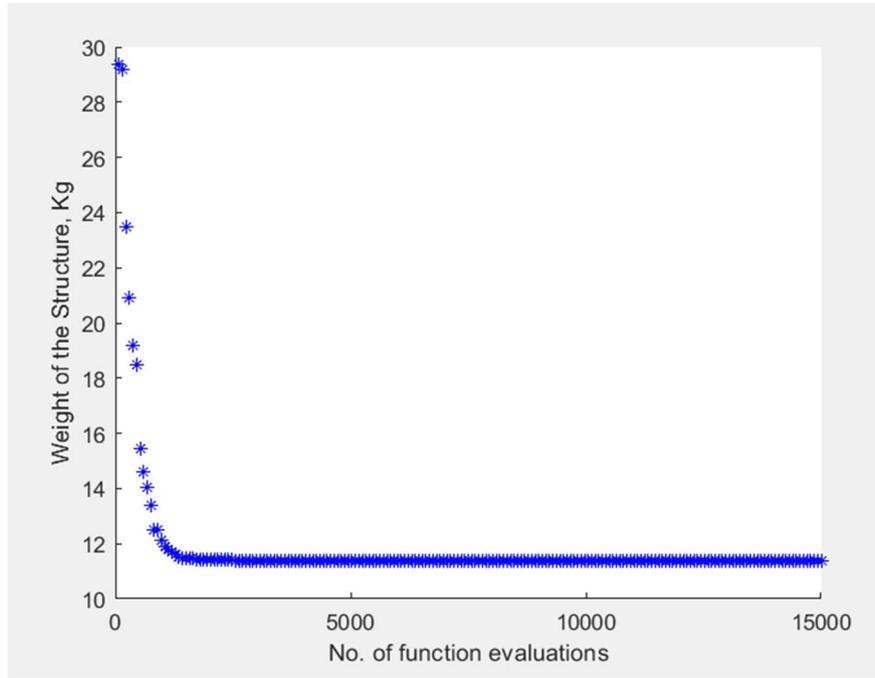

Fig. 20. Convergence graph for the weight optimization on Problem 3

Table 5. Comparison of previously provided area and optimized area for Problem 3

| Cable Member | Previous Area | Optimized Area | Bar Member | Previous Area | Optimized Area |
|---|---|---|---|---|---|
| 1 | 6.28 | 12.57 | 11 | 28 | 145.84 |
| 2 | 6.28 | 12.57 | 12 | 28 | 10.00 |
| 3 | 6.28 | 12.57 | 13 | 28 | 32.20 |
| 4 | 6.28 | 12.57 | 14 | 28 | 66.03 |
| 5 | 6.28 | 12.57 | 15 | 28 | 66.02 |
| 6 | 6.28 | 12.57 | 16 | 28 | 31.98 |
| 7 | 6.28 | 12.57 | 17 | 28 | 10.00 |
| 8 | 6.28 | 12.57 | 18 | 28 | 145.84 |
| 9 | 6.28 | 12.57 | 19 | 28 | 40.74 |
| 10 | 6.28 | 12.57 | 20 | 28 | 39.99 |

## 5.3.2. Optimization on 3D Deployable Structure

Finally, we implement the TLBO algorithm to the deployable 3D structure shown in Problem 4 with population size of 75 and 150 number of iterations. The converged solution for optimized weight is shown in Fig. 21 and Table 6 highlights the member area and forces for some of the members for initial and optimized condition.

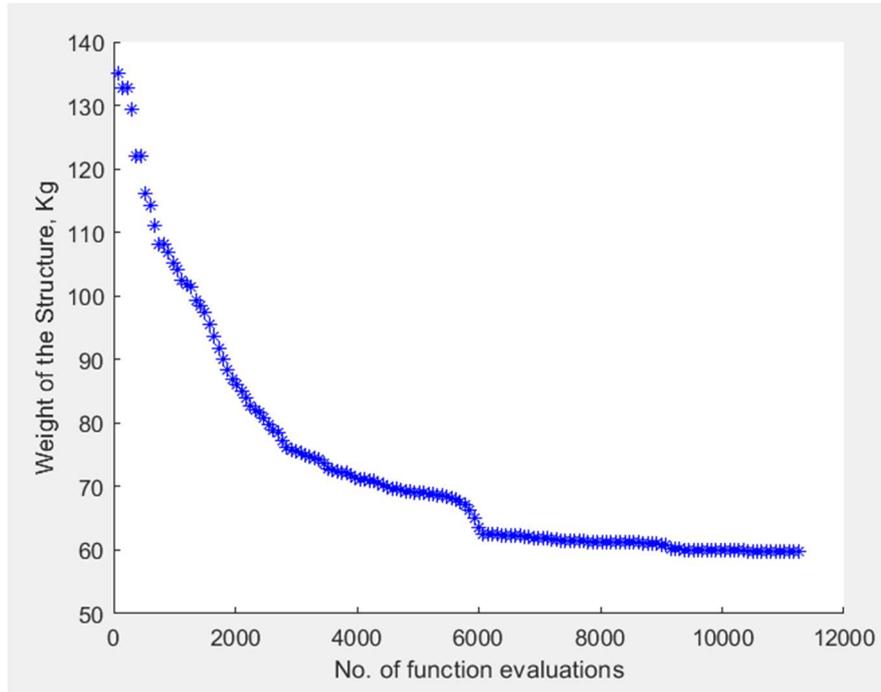

Fig. 21. Convergence graph for the weight optimization on Problem 4

Table 6. Comparison of member force and optimized cross-section area for Problem 4

| Member | Optimized Area | Area Provided | Member Force (Before) | Member Force (After) |
|---|---|---|---|---|
| 1 | 6.28 | 6.28 | 0.00 | 0.00 |
| 2 | 6.28 | 6.28 | -288.22 | -419.78 |
| 3 | 6.28 | 6.28 | -703.17 | -668.23 |
| 10 | 6.28 | 6.28 | -288.22 | -419.82 |
| 11 | 6.28 | 6.28 | 0.00 | 0.00 |
| 12 | 6.28 | 6.28 | -288.22 | -419.75 |
| 13 | 6.28 | 6.28 | -703.17 | -668.22 |
| 33 | 180.73 | 201.06 | 2005.00 | 2101.56 |
| 34 | 43.78 | 50.27 | -462.62 | -420.84 |
| 35 | 54.01 | 63.62 | 310.54 | 314.23 |
| 40 | 183.67 | 201.06 | 2004.98 | 2101.29 |
| 41 | 69.90 | 78.54 | 598.98 | 607.36 |
| 42 | 67.48 | 78.54 | 598.94 | 607.06 |
| 43 | 182.93 | 201.06 | 2005.00 | 2101.59 |
| 62 | 154.45 | 201.06 | 763.35 | 675.46 |
| 63 | 5.00 | 7.07 | 0.00 | 0.00 |
| 64 | 5.00 | 7.07 | 0.00 | 0.00 |

In this scenario, as the size of the structure increases, the cross-sectional area of the members has to be increased significantly (for some cases up to 10 times), to meet the 5 mm allowable displacement requirement. This results in a drastic increase in the weight of the structure. However, it is clear that such an optimization algorithm can give efficient utilization of the material for different configurations. It is important to be noted that such an optimization is based on specific loading scenario. A more general loading case can be defined to locate the most optimum section. However, such analyse is not in the scope of the present work.

## 6. Conclusion

Present manuscript discusses the analysis and optimization of a cable-based deployable pantograph structure using an iterative matrix approach and novel TLBO algorithm. The 3-node uniplet member stiffness is evaluated based on the condensation of the moments at the nodal points. The pantograph structure is a type of mechanism that can expand and contract in a controlled manner, making it useful for various applications such as deployable shelters and bridges. The analysis procedure of cable-based pantograph type of structure based on the algorithm developed by Shan [12], is validated. The same structure has been optimized with a population size of 25 and 100 iterations to achieve the least weight, constrained by 80% of the yield stress of the material (steel) and an upper displacement limit of 5 mm. The optimization results show that all the areas of the members are within a small range, unlike those provided for the validation problem (6.28 mm² for cables and 28 mm² for bars). Similarly, all the member forces are within a small range, unlike those obtained in the validation problem. In the selected structure, evaluation based on a symmetric loading system showed that cable status, reaction forces, member forces, and displacements at relevant nodes are symmetric. The structure also exhibits similar behavior under an asymmetric loading system.

While there are many optimization studies available with truss and scissor-like structures, limited applications have been seen for bar-cable deployable structures. The necessity of cable-strengthened structure is well-versed and helps to increase the stiffness which increases the load bearing capacity and reduces the bending moment. The member buckling is not considered in the analysis and here for the analysis. Further, joint mechanisms are also out of the scope of the present work.

# Appendix A

**Uniplet Stiffness Matrix**

Stiffness matrix, $K_u$ for the uniplet member shown in Figure 2, is as follows:

$$K_u = \begin{bmatrix} k_{11} & 0 & 0 & k_{14} & 0 & 0 & 0 & 0 & 0 \\ 0 & k_{22} & 0 & 0 & k_{25} & 0 & 0 & k_{28} & 0 \\ 0 & 0 & k_{33} & 0 & 0 & k_{36} & 0 & 0 & k_{39} \\ k_{41} & 0 & 0 & k_{44} & 0 & 0 & k_{47} & 0 & 0 \\ 0 & k_{52} & 0 & 0 & k_{55} & 0 & 0 & k_{58} & 0 \\ 0 & 0 & k_{63} & 0 & 0 & k_{66} & 0 & 0 & k_{69} \\ 0 & 0 & 0 & k_{74} & 0 & 0 & k_{77} & 0 & 0 \\ 0 & k_{82} & 0 & 0 & k_{85} & 0 & 0 & k_{88} & 0 \\ 0 & 0 & k_{93} & 0 & 0 & k_{96} & 0 & 0 & k_{99} \end{bmatrix}$$

$$k_{41} = k_{14}, \quad k_{52} = k_{25}, \quad k_{63} = k_{36}, \quad k_{74} = k_{47}$$

$$k_{85} = k_{58}, \quad k_{96} = k_{69}, \quad k_{82} = k_{28}, \quad k_{93} = k_{39}$$

$$m = \frac{1}{L_1}, \quad n = \frac{1}{L_2}$$

$$k_{11} = -k_{41} = E \times A \times m$$

$$k_{22} = 3EI_z m^3 n \Big/ (m+n)$$

$$k_{52} = 3EI_z(-m^3 n - m^2 n^2) \Big/ (m+n)$$

$$k_{82} = 3EI_z m^2 n^2 \Big/ (m+n)$$

$$k_{33} = 3EI_y m^3 n \Big/ (m+n)$$

$$k_{63} = 3EI_y(-m^3 n - m^2 n^2) \Big/ (m+n)$$

$$k_{93} = 3EI_y m^2 n^2 \Big/ (m+n)$$

$$k_{44} = EA(m+n)$$

$$k_{74} = -k_{77} = -E \times A \times n$$

$$k_{55} = 3EI_z(m^3 n + mn^3 + 2m^2 n^2) \Big/ (m+n)$$

$$k_{85} = 3EI_z(-mn^3 - m^2 n^2) \Big/ (m+n)$$

$$k_{66} = 3EI_y(m^3n + mn^3 + 2m^2n^2) \big/ (m+n)$$

$$k_{96} = 3EI_y(-mn^3 - m^2n^2) \big/ (m+n)$$

$$k_{88} = 3EI_z mn^3 \big/ (m+n)$$

$$k_{99} = 3EI_y mn^3 \big/ (m+n)$$